\documentclass[twocolumn,trackchanges]{aastex631}
\usepackage[T1]{fontenc}
\usepackage{amsmath}
\usepackage{orcidlink}
\usepackage{enumitem}
\usepackage{soul}

\newcommand{\e}[1]{\times 10^{#1}}

\newcommand{\tmerger}{t_\mathrm{merger}}

\begin{document}

\title{Black hole pulsars and monster shocks as outcomes of black hole--neutron star mergers}

\author[0000-0002-4305-6026]{Yoonsoo Kim}
\email{ykim7@caltech.edu}
\affiliation{TAPIR, Mailcode 350-17, California Institute of Technology, Pasadena, CA 91125, USA}
\affiliation{Department of Physics, California Institute of Technology, Pasadena, CA 91125, USA}

\author[0000-0002-0491-1210]{Elias R. Most}
\email{emost@caltech.edu}
\affiliation{TAPIR, Mailcode 350-17, California Institute of Technology, Pasadena, CA 91125, USA}
\affiliation{Walter Burke Institute for Theoretical Physics, California Institute of Technology, Pasadena, CA 91125, USA}

\author[0000-0001-5660-3175]{Andrei M. Beloborodov}
\affiliation{Physics Department and Columbia Astrophysics Laboratory, Columbia University, 538 West 120th Street New York, NY 10027,USA}
\affiliation{Max Planck Institute for Astrophysics, Karl-Schwarzschild-Str. 1, 85741 Garching, Germany}

\author[0000-0002-7301-3908]{Bart Ripperda}
\affiliation{Canadian Institute for Theoretical Astrophysics, 60 St. George St, Toronto, ON M5S 3H8, Canada}
\affiliation{Department of Physics, University of Toronto, 60 St. George St, Toronto, ON M5S 1A7, Canada}
\affiliation{David A. Dunlap Department of Astronomy \& Astrophysics, University of Toronto, Toronto, ON M5S 3H4, Canada}
\affiliation{Perimeter Institute for Theoretical Physics, 31 Caroline St. North, Waterloo, ON N2L 2Y5, Canada}

\begin{abstract}
    The merger of a black hole (BH) and a neutron star (NS) {in most cases is
    expected to} leave no material around the remnant BH; therefore, such events
    are often considered as sources of gravitational waves without
    electromagnetic counterparts.  However, a bright counterpart can emerge if
    the NS is strongly magnetized, as its external magnetosphere can experience
    radiative shocks and magnetic reconnection during/after the merger. We use
    magnetohydrodynamic simulations in the dynamical spacetime of a merging
    BH--NS binary to investigate its magnetospheric dynamics. We find that
    compressive waves excited in the magnetosphere develop into monster shocks
    as they propagate outward.
    After swallowing the NS, the BH acquires a magnetosphere that quickly
    evolves into a split monopole configuration and then undergoes an
    exponential decay (balding), enabled by magnetic reconnection and also
    assisted by the ring-down of the remnant BH. This spinning BH drags the
    split monopole into rotation, forming a transient pulsar-like state. It
    emits a striped wind if the swallowed magnetic dipole moment is inclined to
    the spin axis. We predict two types of transients from this scenario: (1) a
    fast radio burst emitted by the shocks as they expand to large radii and (2)
    an X/$\gamma$-ray burst emitted by the $e^\pm$ outflow heated by magnetic
    dissipation.
\end{abstract}

\keywords{Black holes (162), General relativity (641), Gamma-ray bursts (629),
    High energy astrophysics (739), Neutron stars (1108), Plasma astrophysics
    (1261), X-ray bursts (1814), Radio bursts (1339), Transient sources (1851),
    Relativistic binary stars (1386)}

\section{Introduction}

Merging black hole (BH)--neutron star (NS) binaries are promising sources of
gravitational waves (GWs) \cite[see, e.g.][for recent
detections]{LIGOScientific:2021qlt,LIGOScientific:2020zkf,LIGOScientific:2024elc}.
Depending on the mass ratio of the system and spin of the black hole, near-equal
mass systems can feature tidal disruption of the neutron star during merger,
leading to dynamical mass ejection and the formation of a massive disk
\citep{Foucart:2012nc,Foucart:2018rjc}. These can power electromagnetic (EM)
counterparts such as kilonova afterglows
\citep{Lattimer:1974slx,Li:1998bw,Tanaka:2013ixa,Kawaguchi:2016ana,Fernandez:2016sbf,Metzger:2019zeh,Gottlieb:2023vuf,Kawaguchi:2024hdk}
and gamma-ray bursts {(GRBs)}
\citep{Janka:1999qu,Etienne:2011ea,Etienne:2012te,Paschalidis:2014qra,Shapiro:2017cny,Ruiz:2018wah,Hayashi:2021oxy,Gottlieb:2023est,Martineau:2024zur}.
However, the high mass ratio typical of such systems
\citep{LIGOScientific:2021qlt,LIGOScientific:2024elc} would likely result in a
non-disruptive merger, leaving little or no matter surrounding the remnant BH
\citep{Foucart:2012nc,Foucart:2018rjc}. Most BH--NS mergers are expected to fall
in this latter category and be EM-quiet
\citep{Fragione:2021cvv,Biscoveanu:2022iue}, supported by the absence of EM
counterparts to previous detections \cite[e.g.][]{Anand:2020eyg}.

On the other hand, neutron stars can be equipped with strong exterior magnetic
fields, leading to potential EM counterparts from magnetospheric interactions
with their binary companion.
Previously studied scenarios can be broadly split into two groups. Transients
before merger (precursors) can be produced through magnetospheric interactions
\citep{McWilliams:2011zi,Lai:2012qe,Piro:2012rq,Paschalidis:2013jsa,Carrasco:2019aas,Carrasco:2021jja}
including flares \citep{Most:2023unc,Beloborodov:2020ylo}, or through
gravitationally driven resonances in the neutron star such as crustal shattering
\citep{Tsang:2011ad,Penner:2011br,Most:2024qgc}. Potential transients at merger
(concurrent EM counterpart) have been attributed to either a net {electric}
charge of the black hole
\citep{Levin:2018mzg,Zhang:2019dpy,Dai:2019pgx,Pan:2019ulx,Zhong:2019qoi}, or
magnetic flux shedding during the merger process
\citep{DOrazio:2013ngp,Mingarelli:2015bpo,DOrazio:2015jcb,East:2021spd}.

Predicting magnetospheric dynamics of the merger is intrinsically complicated by
various competing processes, some of which can be inferred from previous
numerical studies of a NS gravitationally collapsing into a BH. In this related
scenario, part of the magnetic field is immediately shed during the collapse
\citep{Baumgarte:2002vu,Lehner:2011aa,Palenzuela:2012my,Most:2018abt}. In the
absence of resistive dissipation, the resulting BH can in principle acquire a
net electric charge \citep{Nathanail:2017wly}. However, pair-production in
realistic environments will lead to an active magnetosphere supporting magnetic
flux decay (balding) of the BH
\citep{Lyutikov:2011tk,Bransgrove:2021heo,Selvi:2024lsh}.
On a technical level, most of the studies in numerical relativity have made use
of the force-free electrodynamics or vacuum approaches to study magnetospheric
dynamics. Compared to magnetohydrodynamic (MHD) approaches explicitly tracking
matter dynamics, these crucially miss out the formation of monster radiative
shocks from fast magnetosonic waves \citep{Beloborodov:2022pvn} as was recently
demonstrated by \cite{Most:2024qgc}, which could be responsible for some of the
high-energy emission in this process.

Here, we present general relativistic (GR-)MHD simulations in full numerical
relativity of a merging BH\textendash{}NS binary, in which the NS is swallowed
whole. While BH--NS merger simulations in GRMHD have become common (e.g.,
\citealt{Chawla:2010sw,Etienne:2011ea,Etienne:2012te,Kiuchi:2015qua,Ruiz:2018wah,Ruiz:2020elr,Most:2021ytn,Hayashi:2021oxy,Hayashi:2022cdq,Izquierdo:2024rbb},
tracking the magnetospheric evolution requires special flooring techniques
\citep{Tchekhovskoy:2012hm,Parfrey:2017nby}. We employ such a sophisticated MHD
strategy to track the evolution of magnetosphere throughout inspiral and merger.
Our simulations identify novel types of shock-powered and reconnection-driven
transients from a BH--NS merger. Specifically, we show that monster shocks are
formed during the final phase of the inspiral, which can primarily source X-ray
and radio bursts.
In the post-merger phase, we find that the magnetosphere of the {remnant} BH
re-arranges into a short-lived {\it black hole pulsar} state
\citep{Selvi:2024lsh}, capable of powering X-ray transients that may last for
several milliseconds.

We describe the simulation setup and the configuration of the binary in
Sec.~\ref{sec:methods}. This is followed by detailed discussions of two new
transients from {non-disrupting} BH--NS mergers. First, we present the formation
of monster shocks in Sec.~\ref{sec:monster shock}. Next, we provide a detailed
analysis of the black hole pulsar state that our simulations reveal in
Sec.~\ref{sec:bh pulsar}. {We discuss the properties of the expected EM
emissions in Sec.~\ref{sec:em transient}.} Finally, we conclude by summarizing
our findings in Sec.~\ref{sec:conclusion}. Unless otherwise stated, we adopt
Gaussian units with $c=G=1$ throughout this paper.

\section{Methods} \label{sec:methods}

\begin{figure*}
    \centering
    \includegraphics[width=\linewidth]{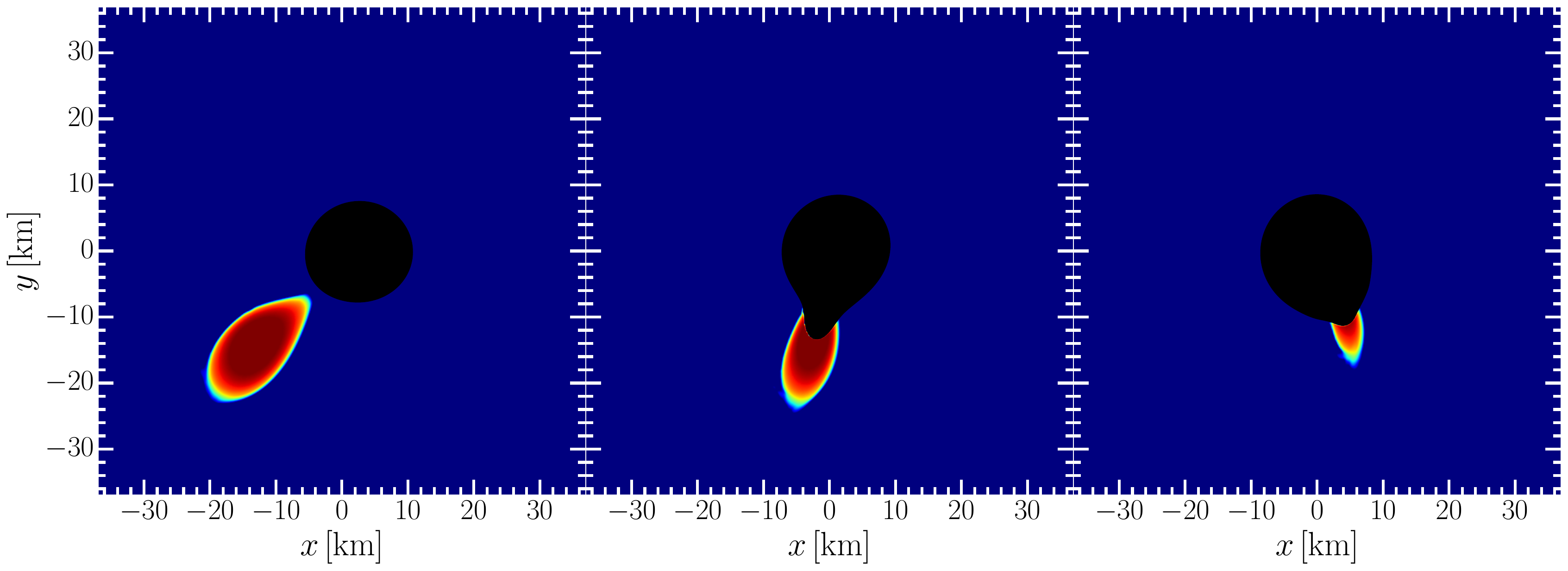}
    \caption{Merger of the BH--NS binary in our simulations, where the neutron
    star is swallowed whole. The entire process {shown in this figure} happens
    in less than one millisecond.}
    \label{fig:merger}
\end{figure*}

We track the time evolution of a BH--NS binary as well as the common binary
magnetosphere using ideal GRMHD for dynamical spacetimes \citep{Duez:2005sf}. To
this end, we need to specify both initial conditions and evolution parameters.

We use the \texttt{Kadath/FUKA} \citep{Papenfort:2021hod,Grandclement:2009ju}
initial data framework to construct BH--NS initial data in extended conformal
thin sandwich (XCTS) form
\citep{Grandclement:2006ht,Taniguchi:2007xm,Taniguchi:2007aq,Foucart:2008qt,Tacik:2016zal}.
In order to ensure that the NS is fully swallowed at merger,\footnote{See
\cite{Foucart:2012nc,Foucart:2018rjc} for the allowed parameter space of a
non-disrupting BH--NS merger.} we adopt a non-spinning {NS} with mass
$M_\text{NS} = 1.4M_\odot$ using the APR4 equation of state
\citep{Akmal:1998cf}, and a BH with mass $M_\text{BH} = 8.0M_\odot$ and
dimensionless spin $a=0.3$ aligned with the orbital axis ($\hat{z}$). The
initial orbital separation of the binary is 60km, resulting in $\sim 1.5$ orbits
before the merger.
The neutron star is initially magnetized with a dipolar field with a strength
$|B_*| = 1.9\e{16} \,{\rm G}$ at the magnetic poles on the surface. The precise
value of the magnetic field is unimportant for the magnetospheric dynamics we
study, {since we fix the properties of the magnetosphere in terms of
dimensionless quantities such as magnetization $\sigma = b^2/\rho$, and plasma
$\beta = 2 P/b^2$, where $b^2$ is the magnetic energy density, $\rho$ the
rest-mass density and $P$ the pressure. This allows us to rescale the resulting
magnetospheric dynamics to arbitrary magnetic field strength.} However, for
purely numerical reasons we have found that using a stronger field strength
eases the transition to a near force-free magnetosphere near the stellar surface in the inspiral computationally.
Nevertheless, the chosen strength of the magnetic field hardly impacts the bulk
dynamics of the NS (the plasma beta parameter is around $\beta \sim 10^3$ inside
the NS during the inspiral).
We simulate three models with {an initial inclination between the magnetic
dipole moment and the orbital axis} $\theta_B=0^{\circ}$, $30^{\circ}$, and
$60^{\circ}$. The initial NS magnetic field is inclined toward the companion
{BH} at $t=0$.

Dynamical evolutions are performed with the \texttt{Einstein Toolkit} framework
\citep{Loffler:2011ay}, using the \texttt{Frankfurt/IllinoisGRMHD (FIL)}
\citep{Most:2019kfe,Etienne:2015cea} code for solving the ideal GRMHD equations
in a dynamical spacetime. The spacetime is evolved using \texttt{FIL}'s
numerical relativity solver, which implements the Z4c equations
\citep{Bernuzzi:2009ex,Hilditch:2012fp} in moving puncture gauge
\citep{Alcubierre:2002kk} using a fourth-order finite-difference discretization
\citep{Zlochower:2005bj}. The ideal GRMHD equations are solved using the ECHO
scheme \citep{DelZanna:2007pk} with upwind constraint transport
\citep{Londrillo:2003qi}. Similar to our previous work \citep{Most:2024qgc}, the
fourth-order derivative corrector in the ECHO scheme showed less robust behavior
at strongly magnetized shockfronts, and we have disabled it in {our} runs. A key
feature of our simulations is the ability to track the common magnetospheric
dynamics in full MHD as opposed to vacuum or force-free electrodynamics. While
several studies have evolved magnetic fields in the exterior region in the
context of BH--NS mergers \citep{Paschalidis:2014qra,Ruiz:2018wah,Ruiz:2020elr},
reproducing correct (near-) force-free magnetospheric dynamics {within the MHD
formulation} requires the use {of robust primitive inversion schemes
\citep{Kastaun:2020uxr}} and special flooring techniques
\citep{Tchekhovskoy:2012hm,Parfrey:2017nby}, unlike floors commonly used in
numerical relativity simulations \cite[e.g.][]{Poudel:2020fte}. A detailed
prescription of the floors we use here is provided in \citet{Most:2024qgc}. It
is precisely this flooring scheme that allows us to correctly capture and
uncover the transients we present in this study. Similar to
\citet{Most:2024qgc}, we have supplemented the high-density cold equation of
state used in the initial data with a thermal equation of state, $P_{\rm th} =
\rho \epsilon$, which primarily governs the magnetospheric dynamics. Here
$P_{\rm th}$ is the thermal pressure, and $\epsilon$ the specific {internal}
energy. We use a three-dimensional Cartesian grid with eight levels of nested
moving mesh refinement \citep{Schnetter:2003rb}. The coarsest grid extends to
$[-3025 {\rm km}, 3025 {\rm km}]^3$ and the finest resolution is $168\, \rm m$.
The finest grid level consists of two patches covering 2--3$\times$ the size of
the NS as well as of the BH, being centered to and tracking each of them.

A detailed description of the initial evolution of non-disruptive {BH--NS}
mergers can be found elsewhere \cite[see e.g.][for a recent
review]{Kyutoku:2021icp}. Since the magnetospheric transients in our simulations
are mainly driven during and after the merger, we briefly depict the merger
process in Figure~\ref{fig:merger}, which highlights the high degree of spatial
asymmetry present in the process, and consequently, the need for full numerical
relativity not only for the spacetime evolution but particularly to correctly
determine the geometry of magnetic field in the post-merger phase.

\section{Monster shock} \label{sec:monster shock}

\begin{figure*} 
    \centering
    \includegraphics[width=\linewidth]{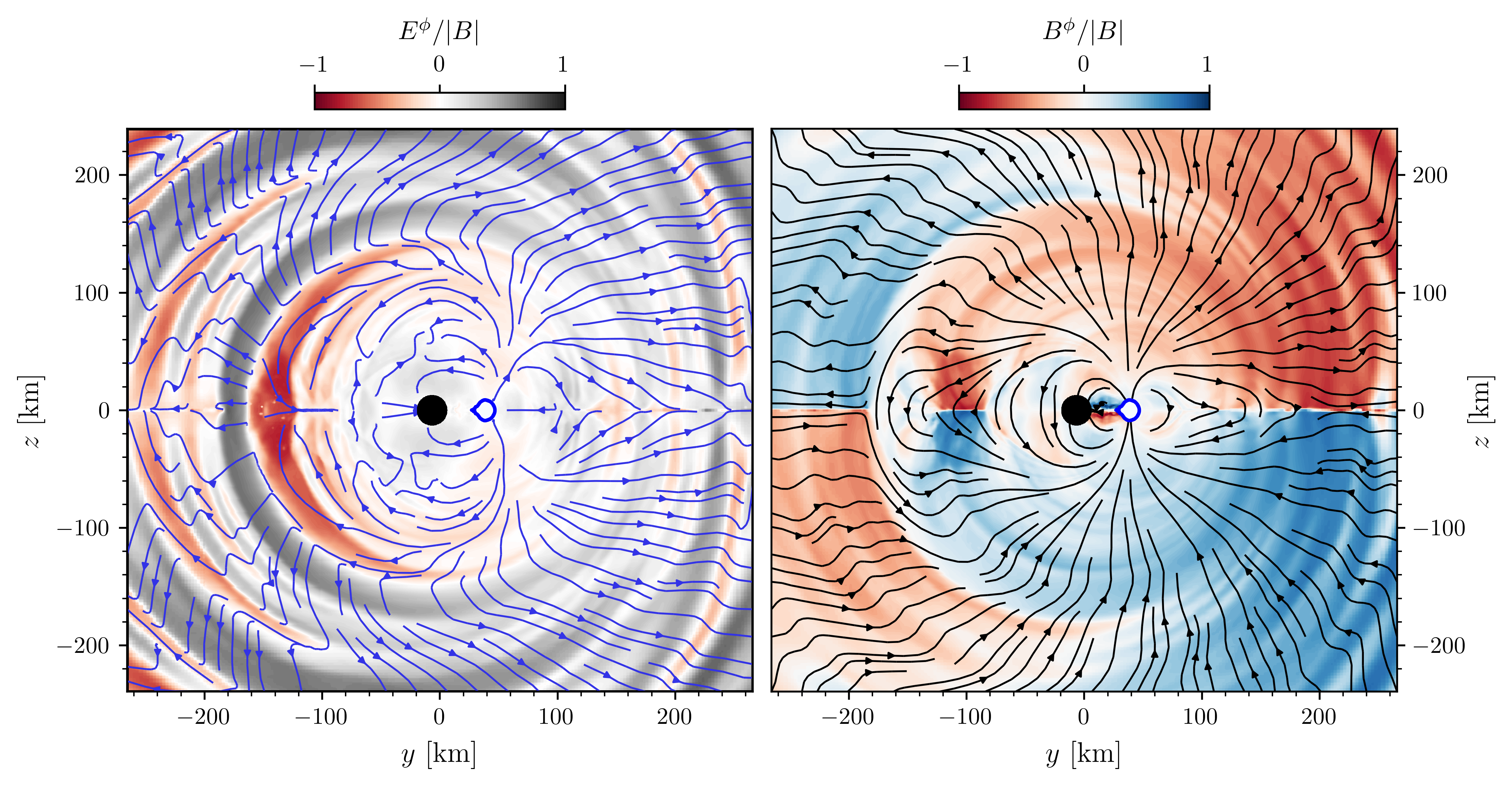}
    \caption{ Poloidal structure (cut in the $yz$ plane) of the perturbed
    magnetosphere of the BH--NS binary 0.9 ms before merger for the aligned
    ($\theta_B = 0^\circ$) model. Fast magnetosonic waves have toroidal electric
    fields $E^\phi$ (left), and Alfv\'en waves have toroidal magnetic
    perturbations $\delta B^\phi$ (right). Streamlines show fluid velocity in
    the left panel and magnetic field lines in the right panel. The BH and NS
    are shown with a black and blue circle, respectively.}
    \label{fig:pre merger magnetosphere}
\end{figure*}

\begin{figure*}
    \centering
    \includegraphics[width=\linewidth]{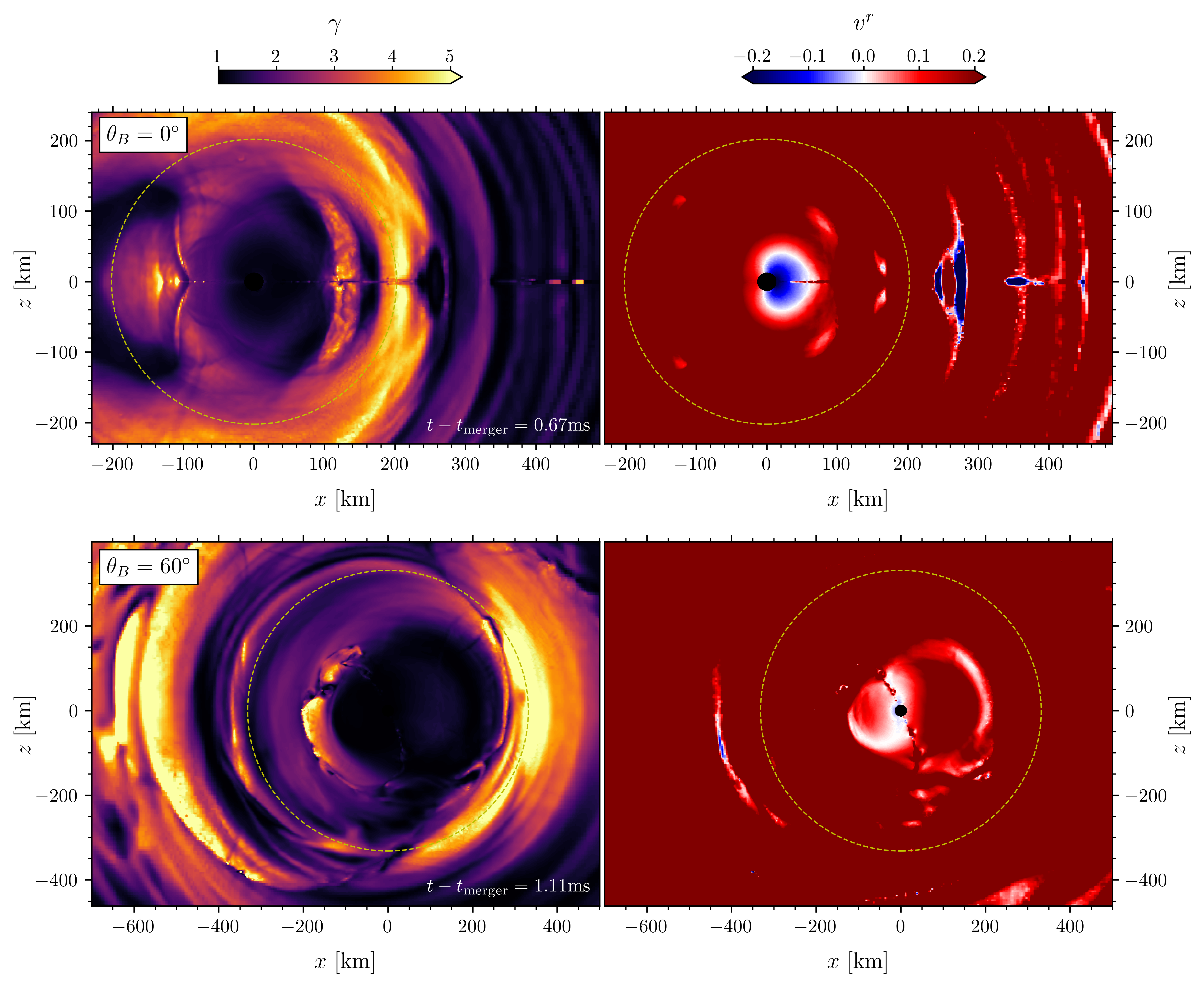}
    \caption{Monster shocks launched from BH--NS mergers. Shown here are the
        Lorentz factor (left panels) and the radial spatial velocity (right
        panels) on the meridional ($xz$) plane. Dashed orange circles are light
        spheres with the radius $r = c(t-\tmerger)$.
        (Top) Simulation snapshot from the aligned model ($\theta_B = 0^\circ$).
        A monster shock can be found near $x \simeq 250{\rm km}$, with its
        characteristic feature of a plasma moving radially inward ($v^r < 0$)
        preceding the shockfront. (Bottom) Inclined model $\theta_B = 60^\circ$.
        A similar feature can be seen near $x\simeq -400 {\rm km}$.}
    \label{fig:ms}
\end{figure*}

The initially dipolar magnetosphere of the NS is sheared in the vicinity of the
BH before and during merger. This perturbation will launch waves into the
magnetosphere, which will either be transverse (Alfv\'en) wave propagating along
the magnetic field, or be a compressional (fast magnetosonic) wave. In a dipole
background field, the compressional waves are expected from any non-toroidal
perturbation in the magnetosphere, as happens during the merger.

Propagation of fast magnetosonic waves to larger radii $r$ is not affected by
the background field as long as their {wave} amplitude $E=\delta B\propto
r^{-1}$ is much smaller than the background dipole field $B_{\rm bg}\propto
r^{-3}$. With increasing $r$, this condition becomes broken, $B^2-E^2$
approaches zero, and the plasma drift speed in the wave approaches the speed of
light.\footnote{Orbiting systems possess an orbital light cylinder, $r_{\rm LC}
\sim 1/\Omega_{\rm orb}$, set by the orbital frequency $\Omega_{\rm orb}$.
Steepening or distortion of the waves induced by a decreasing $B_{\rm bg}$} will
only happen efficiently on closed field lines inside the light cylinder, in turn
requiring a minimum amplitude of the perturbation (see, e.g.,
\citet{Most:2024eig} for a discussion in the context of non-linear steepening of
Alfv\'en waves).
Recent analytical \citep{Beloborodov:2022pvn} and numerical
\citep{Chen:2022yci,Vanthieghem:2024van} works have demonstrated that this leads
to the formation of {\it monster shocks} (see also \citet{Lyubarsky:2002bw} for
earlier work). In particular, in the equatorial plane of the magnetic dipole,
the shock appears when $\delta B \approx B_{\rm bg}/2$, which implies $B^2-E^2$
touching zero at the trough of the compressional wave. Near this point, the
plasma develops a characteristic negative velocity $v^r<0$, which leads to shock
formation in front of the crest of the wave. In practice, searching for zones
with $v_r<0$ provides a simple way to identify regions of shock formation, in
addition to detection of velocity jumps and localized heating spikes. A similar
analysis was performed in \citet{Most:2024qgc} to demonstrate shock formation in
the magnetosphere of a collapsing magnetar.

In our simulations, the inspiral of the magnetized NS drives a continuous
excitation of magnetosonic waves in the magnetosphere, peaking around the plunge
of the NS into the BH.
The final plunge of the NS injects a strong rarefaction mode into the
surrounding magnetosphere as the NS bulk velocity is maximally radially inward
at the moment.
In Fig. \ref{fig:pre merger magnetosphere}, we show the excited magnetosphere
about half an orbit before the plunge for aligned ($\theta_B = 0^\circ$)
magnetic axis.
We find that the wave emitted during the plunge leads to the development of a
large $v^r < 0$ region characteristic of the monster shock, which we show in the
top row of Fig.~\ref{fig:ms}. This phenomenology of a leading shock with
surrounding weaker shocks resembles the results for the collapsing magnetar
\citep{Most:2024qgc},
and approximately agrees with the analytical prediction \cite[][see Fig. 7
therein]{Beloborodov:2022pvn}.

The profile of $\gamma v^r$ across the shock 
region is affected by deviations of $B_{\rm bg}$ from a pure dipole due to the
orbital motion of the NS. As an additional validation, we have also confirmed
that the regions with $v^r < 0$ develop $E^2\approx B^2$ plateaus, corroborating
that the observed feature is the monster shock.

We have also identified monster shocks for the inclined models. One such model
with $\theta_B = 60^\circ$ is shown in in the bottom row of Fig.~\ref{fig:ms}.
We caution that due to the misalignment between magnetic equator and orbital
plane, the strongest part of the shock will appear off the shown $yz$ plane, 
and the trough of the wave preceding the monster shock might not strongly
exhibit $v^r < 0$. Yet, we can clearly identify a similar leading shock
structure as in the aligned case.

\section{Transient black hole pulsar}
\label{sec:bh pulsar}

In this section, we present a detailed analysis of the evolution of the
post-merger magnetosphere, defined as the region within the light sphere ($r/c
\leq t - \tmerger$), with a special emphasis on the near-horizon dynamics. The
merger remnant settles down to a Kerr BH with the mass $M = 9.2M_\odot$ and the
dimensionless spin $a = 0.57$. Relevant length and time scales are $r_g \equiv
GM/c^2 = 13.6 {\rm km}$ and $r_g/c = 46\, {\rm \mu s}$,
or equivalently 1 millisecond amounts to $\sim 22 r_g/c$. The angular velocity
of the outer event horizon is $\Omega_H = ac/2r_+ = 3.43\times 10^3 {\rm
s^{-1}}$,
where $r_+ = r_g(1 + \sqrt{1-a^2})\approx 25\, \rm km$ is the outer horizon
radius. While we will quote values measured from our simulation data in the
following discussions, we caution that {it is nontrivial to map} our results
(especially time scales) in a coordinate-independent manner to those obtained
from other studies that used a fixed Kerr background, since our merger
simulations are performed using dynamically evolved coordinates.

\subsection{Relaxation into a rotating split-monopole}
\label{sec:split monopole formation} 

\begin{figure*}
    \centering
    \includegraphics[width=\linewidth]{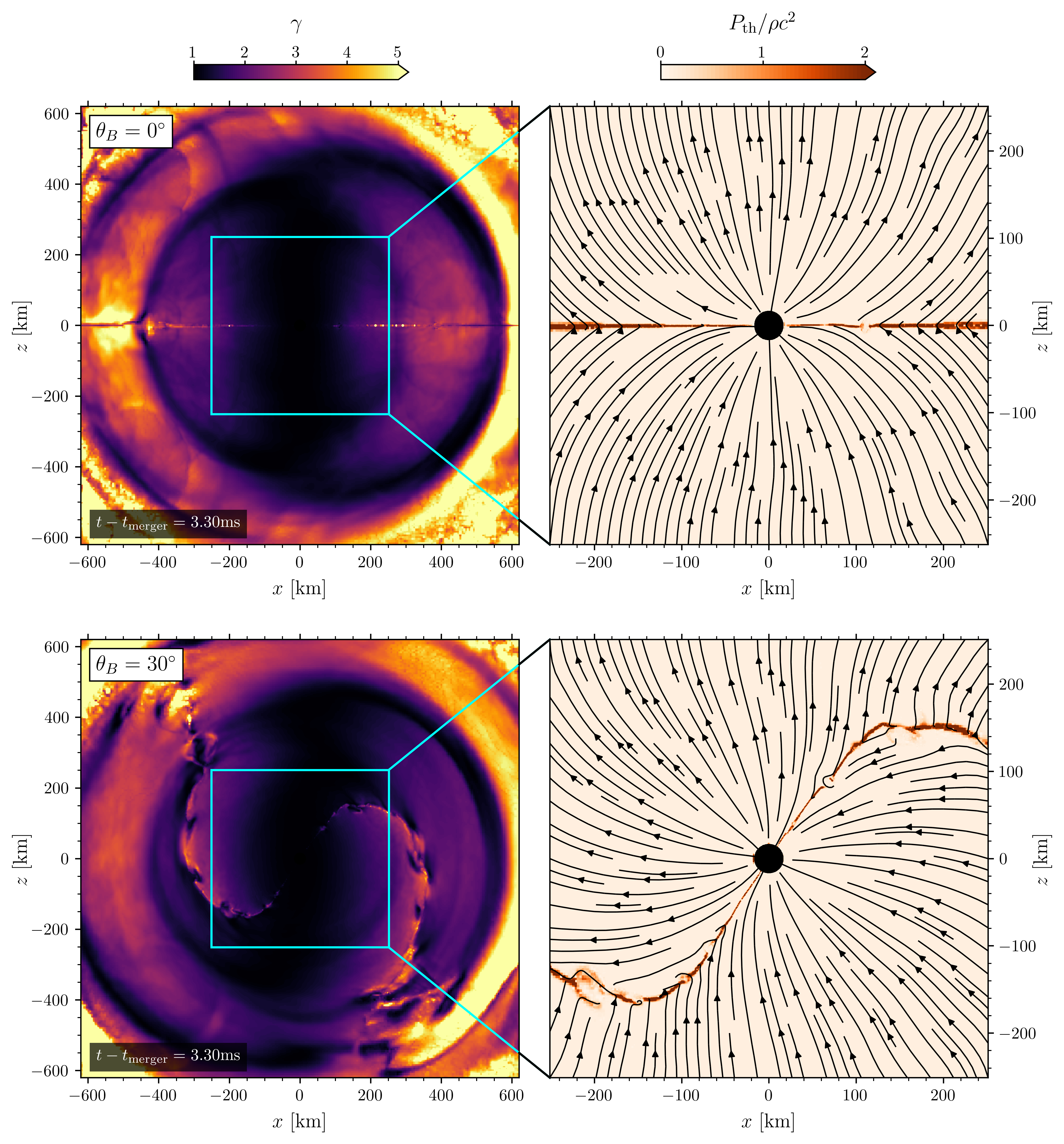}
    \caption{Post-merger magnetosphere of the remnant black hole having settled
        down to a rotating split monopole. {The physical quantities are shown}
        in the $xz$ plane. Left: fluid Lorentz factor $\gamma$. Right: 
        ratio of the thermal pressure $p_\mathrm{th}$ to the rest energy density
        $\rho c^2$. Black solid lines show the in-plane magnetic field lines. An
        equatorial current sheet is formed at which the magnetic field lines in
        upper and lower hemispheres reconnect, dissipating magnetic energy and
        causing a flux decay (balding) of the BH. The inner magnetosphere is
        driven to co-rotate with the BH due to frame dragging. As a result, the
        post-merger magnetosphere of an inclined model (e.g.
        $\theta_B=30^\circ$, bottom panels) exhibits features similar to those
        of tilted pulsars (see also Fig.~\ref{fig:striped wind}). The spin axis
        of the BH is along $\hat{z}$ in all models.
        }
    \label{fig:bhpulsar}
\end{figure*}

\begin{figure}
    \centering
    \includegraphics[width=\linewidth]{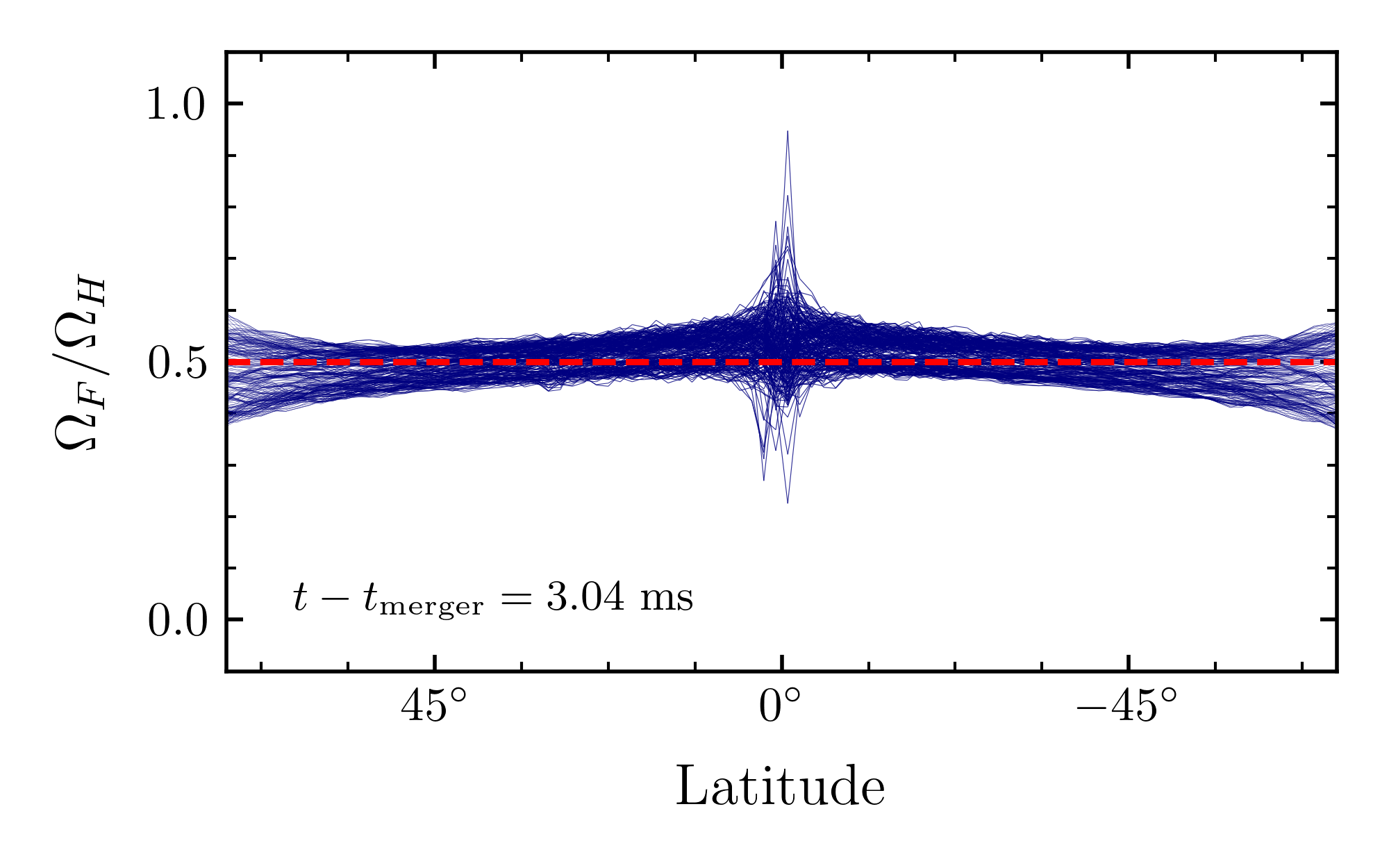}
    \caption{{Angular velocity of magnetic field lines threading the apparent BH
        horizon for $\theta_B=0^\circ$ simulation. Shown are the distribution of
        $\Omega_F$ for each latitude
        and the stationary axisymmetric force-free solution $\Omega_F \simeq
        \Omega_H/2$ (red dashed line).}}
    \label{fig:magnetic field angular velocity}
\end{figure}

The remnant BH is immersed in a dipole-like magnetic field shortly after the
merger. Since black holes cannot support closed magnetic field lines
\citep{MacDonald:1982zz}, the dipole gets stretched out, with the magnetic field
lines opening up near the magnetic equator. In consequence, the BH
{magnetosphere} transitions into a split-monopole topology
\citep{Komissarov:2004qu}, and begins to dissipate the magnetic field energy at
the current sheet. The inclination of the split-monopole configuration depends
on the initial inclination of the NS magnetic field. For all simulations, the
topology of magnetic field lines transitions into a split-monopole over a
timescale of 
1\,ms,\footnote{A similar reordering and collimation of the post-merger magnetic
field may also have been observed in previous works \citep{East:2021spd}.} which
is consistent with multiple light crossing times across the horizon ($2r_+/c
\approx\,  0.2\, {\rm ms}$).

The distribution of magnetic flux on the remnant BH is initially highly
localized to the spot through which the NS plunged into (Fig. \ref{fig:merger}).
Over the transition period to a split-monopole ($t-\tmerger \lesssim {\rm 1
ms}$), the magnetic flux density on the BH horizon is redistributed, relaxing
into a relatively uniform distribution by $t-\tmerger\approx {\rm 2\, ms}$. The
upper panels of Fig.~\ref{fig:bhpulsar} show the post-merger magnetosphere at
$t-\tmerger = 3.3\, \text{ms}$ for the aligned model ($\theta_B = 0^\circ$),
displaying an axisymmetric split-monopole magnetosphere centered on the BH.
Magnetic energy of the magnetosphere is partially dissipated via reconnection in
the equatorial current sheet, heating the plasma, as can be seen from 
the plot of $P_{\rm th}/\rho$ in Fig.~\ref{fig:bhpulsar}.

The frame dragging of the remnant BH induces co-rotation of magnetic field lines
and forms a rotating split-monopole. The angular velocity of the magnetic field
lines in an axisymmetric force-free split-monopole magnetosphere around a Kerr
BH is given as $\Omega_F = a/8M$ to leading order in {the spin}
\citep{Komissarov:2004ms,Armas:2020mio}.
For arbitrary high spins, $\Omega_F$ can be calculated either with a
perturbative analytic expansion \cite[e.g.][]{Armas:2020mio} or using an
iterative numerical method \cite[e.g.][]{Contopoulos:2012py,Nathanail:2014aua}.
The ratio $\Omega_F / \Omega_H$, which is $1/2$ in the limit $a \to 0$, remains
$\lesssim 1 \%$ different from $1/2$ for the spin $a \leq 0.7$ \cite[e.g. see
Figure 1 of][]{Contopoulos:2012py}. Therefore, in the case of our merger remnant
BH with $a=0.57$, we can safely assume $\Omega_F / \Omega_H \simeq 0.5$.

For the aligned ($\theta_B = 0^\circ$) model, we measure the rotation angular
velocity of magnetic field lines as
\begin{equation} \label{eq:omega-f}
    \Omega_F = \frac{-y \,(u^x/u^0) + x \,(u^y/u^0)}{\varpi^2},
\end{equation}
where $u^\mu$ is the four-velocity of the plasma and $\varpi = \sqrt{r^2-z^2}$
is the (coordinate) cylindrical radius. This description is appropriate for the
ideal MHD limit we consider. Fig.~\ref{fig:magnetic field angular velocity}
shows the measured $\Omega_F/\Omega_H$ over a spherical surface $r=2.4\, r_g$
encompassing the remnant BH.\footnote{When computing the horizon angular
velocity $\Omega_H$, we used the mean radius of the instantaneous apparent
horizon of the BH.}
We observe that the rotation angular velocity of magnetic field lines converges
to $\Omega_H/2$ and the asymmetry present in its distribution is decayed out
over time. In addition to the magnetic field morphology shown in
Fig.~\ref{fig:bhpulsar}, this provides solid evidence that the post-merger
magnetosphere relaxes into a rotating split-monopole.

As naturally expected, for inclined magnetic field, the remnant BH settles down
to a rotating, inclined split-monopole magnetosphere. The resulting global
dynamics of the magnetosphere closely resembles that of a tilted pulsar, akin to
a recently proposed {\it black hole pulsar} state \citep{Selvi:2024lsh}. We
present detailed discussions on this transient BH pulsar in later sections.

\subsection{Rotation and alignment of current sheets}
\label{sec:cs rotation and alignment}

\begin{figure*}
    \centering
    \includegraphics[width=\linewidth]{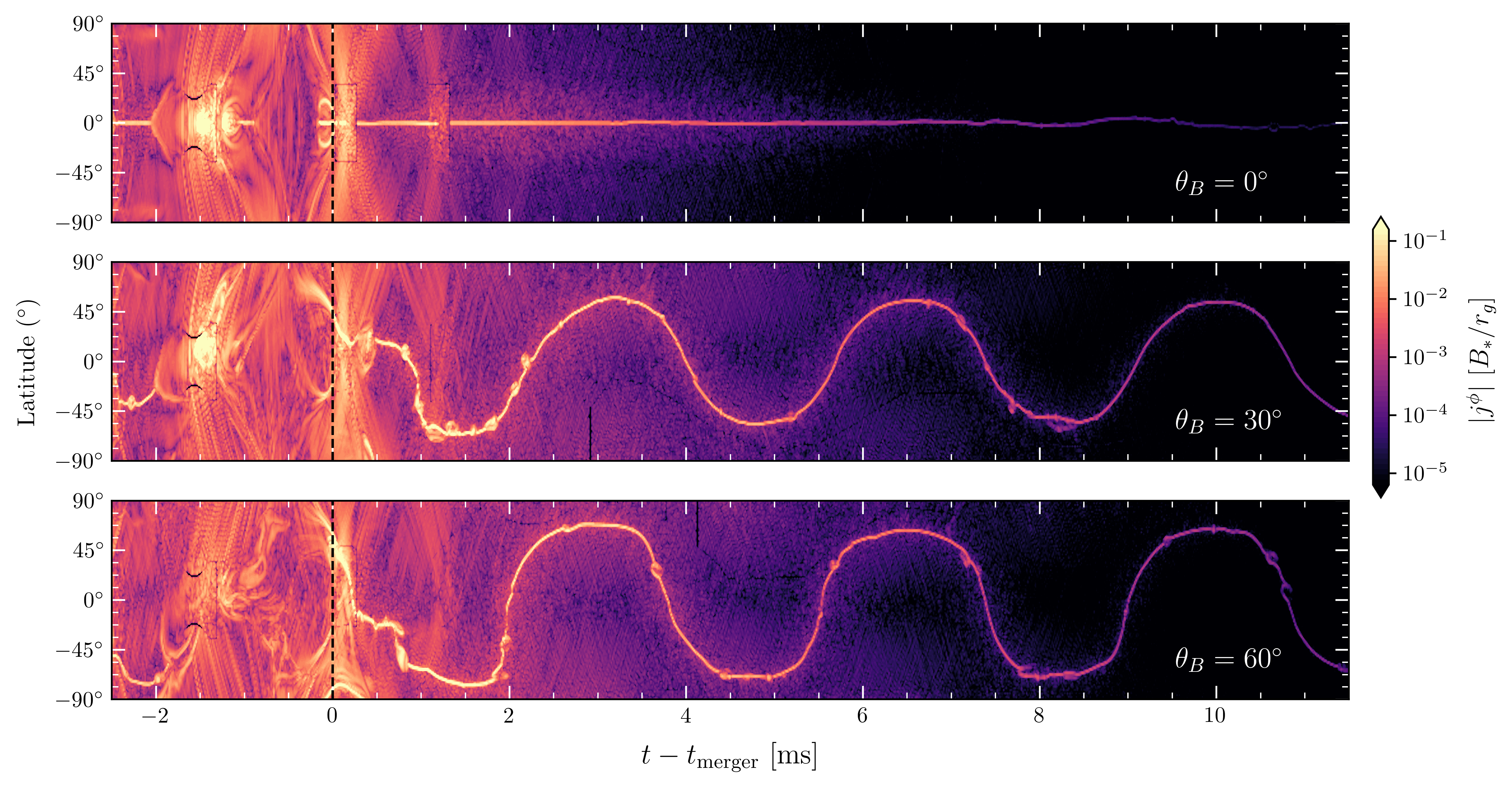}
    \caption{A spacetime diagram of the approximate electric current
        $\left|j^\phi\right|$ on a meridional arc $\phi=0$ at $r=2.4r_g$ 
        {in three simulations with different {initial} inclinations of the {NS}
        magnetic dipole moment ($\theta_B =0^\circ, 30^\circ, 60^\circ$),}
        displaying the latitude of the {post-merger} magnetospheric BH current
        sheet. For the inclined models ($\theta_B = 30^\circ, 60^\circ$), the
        periodic oscillation in the latitude represents the rotation of the
        inclined current sheet. The orbital current sheet of the NS in the
        inspiral phase is also visible for $t-\tmerger \lesssim -2\, {\rm ms}$,
        where $\tmerger$ indicates the merger time.}
    \label{fig:cs}
\end{figure*}

\begin{figure}
    \centering
    \includegraphics[width=\linewidth]{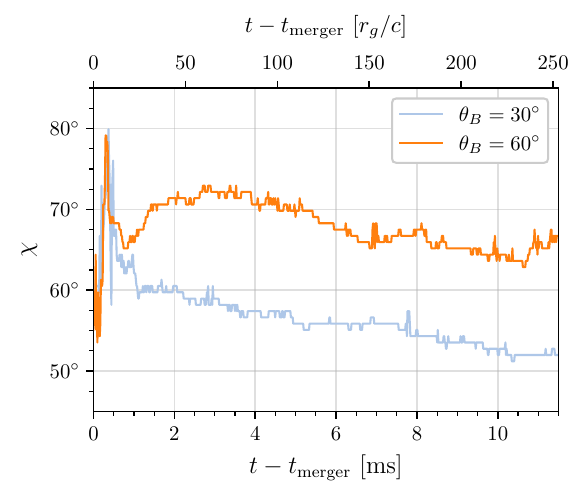}
    \caption{Time evolution of the current sheet inclination angle $\chi$ for
        the $\theta_B = 30^\circ, 60^\circ$ models.}
    \label{fig:cs alignment}
\end{figure}

The spinning remnant BH induces a rotation of the magnetic field lines and the
current sheets via frame dragging with respect to its spin axis. In the
following, we would like to track the motion of current sheets. These are easily
identified with the (toroidal) electric current, $j^\phi$, which we here
approximate via its Newtonian expression,
\begin{equation} \label{eq:jphi}
    j^\phi \approx \epsilon^{\phi i j} \partial_i B_j\,.
\end{equation}
We then analyze the time evolution of the current on a fixed spherical surface
of radius $r=2.4r_g$. In Fig.~\ref{fig:cs}, we show the distribution of
$|j^\phi|$ on the meridional arc $\phi=0$ ($x>0, y=0$) for all simulations. The
equatorial current sheet in the aligned case ($\theta_B=0^\circ$) does not
exhibit notable modulations in its latitude, where we have confirmed in
Sec.~\ref{sec:split monopole formation} that the magnetosphere is in fact
rotating with $\Omega_F = 0.5 \Omega_H$. For the inclined cases, rotation of the
current sheet is clearly seen in Fig.~\ref{fig:cs}. The time interval between
neighboring peaks is about 3.5\,ms, revealing that the current sheet is rotating
with about half {of} the horizon angular velocity.

Over each meridional lines ($\phi = {\rm const.}$) on the spherical surface
$r=2.4r_g$, we collect the latitude $\alpha_0(\phi)$ with the maximum value of
$|j^\phi|$, which is effectively the latitude of the current sheet at that
azimuthal angle. Then we define the current sheet inclination angle $\chi$
as\footnote{See the section 3 of \cite{Selvi:2024lsh} for an alternative method
to measure $\chi(t)$ in terms of the magnetic moment.}
\begin{equation} \label{eq:cs inclination angle}
    \chi = \frac{ \max[\alpha_0(\phi)] - \min [\alpha_0(\phi)] }{2} .
\end{equation}
In Fig.~\ref{fig:cs alignment}, we show the measured $\chi(t)$ from $\theta_B =
30^\circ, 60^\circ$ simulations. A nonlinear deformation of the NS during the
merger is found to greatly enhance the inclination angle of the magnetic field
around the merger remnant, resulting in $\chi\approx 60^\circ$ for $\theta_B =
30^\circ$. We observe a gradual decay in $\chi(t)$ for both models, indicating
the alignment of the current sheet with respect to the BH spin axis over time,
which is consistent with the result of \cite{Selvi:2024lsh}. However, here we
can only provide a crude estimate on the alignment timescale $\tau_\chi \approx$
1000\textendash{}2000\,$r_g/c$, being limited by a short simulation time and a
mild spin of the BH. We also caution that the observed alignment timescale could
be affected by a high numerical dissipation (see Sec. \ref{sec:balding and
ringdown}).

\subsection{Balding and ring-down of the remnant BH}
\label{sec:balding and ringdown}

\begin{figure}
    \centering
    \includegraphics[width=\linewidth]{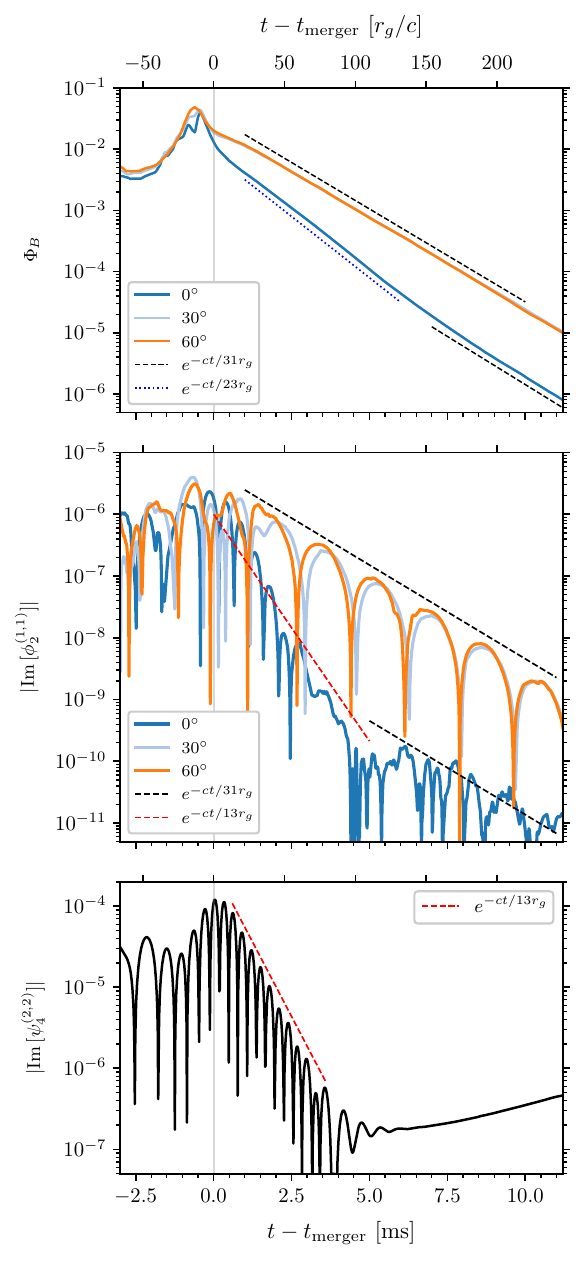}
    \caption{ Top: total magnetic flux extracted on a spherical surface
        $r=2.4r_g$ near the apparent horizon. The result from $\theta_B =
        30^\circ$ (cyan solid line) and $\theta_B = 60^\circ$ (orange solid
        line) are lying almost on top of each other.
        Middle: imaginary part of $(l,m)=(1,1)$ mode of the Maxwell
        Newman-Penrose (NP) scalar $\phi_2$ extracted at $r=4.3r_g$.
        Bottom: imaginary part of $(l,m)=(2,2)$ mode of the NP scalar $\psi_4$
        extracted at $r=4.3r_g$. We only show the result from $\theta_B =
        0^\circ$ since the result is almost identical for all simulations. 
        Exponential decays
        with timescales $31r_g/c$, $23r_g/c$ and $13r_g/c$ are indicated by the
        black dashed, blue dotted, and red dashed lines.
        }
    \label{fig:balding}
\end{figure}

\begin{figure*}
    \centering
    \includegraphics[width=\linewidth]{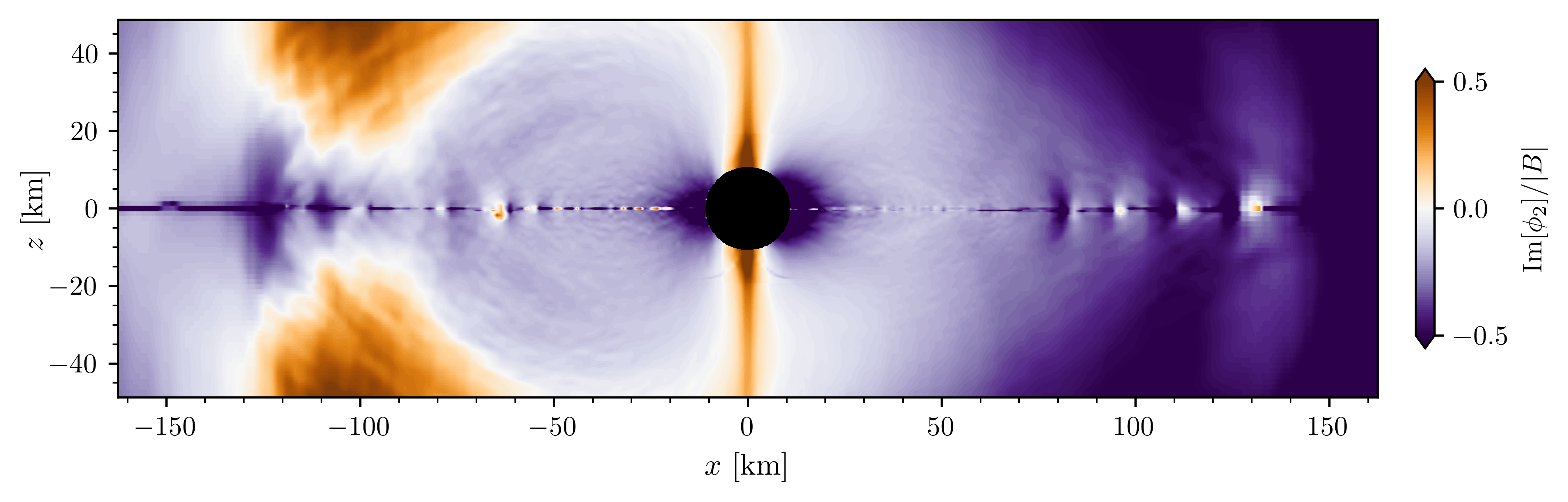}
    \caption{Imaginary part of the Maxwell Newman-Penrose scalar $\phi_2^{(l=1,
        m=1)}$, corresponding approximately to outgoing fast magnetosonic waves,
        on the vertical ($xz$) plane normalized with the magnitude of magnetic
        field for the aligned case $\theta_B = 0^\circ$ at $t-\tmerger =
        1.05\,{\rm ms}$.}
    \label{fig:ringdown magnetosphere}
\end{figure*}

In a split-monopole magnetosphere of a stationary BH, the total magnetic flux
threading the horizon
\begin{equation}
    \Phi_{\rm B} = \frac{1}{2} \oint |B^r| d\Omega , 
\end{equation}
exponentially decays as a result of magnetic reconnection in the current sheet
\citep{Lyutikov:2011tk,Bransgrove:2021heo,Selvi:2024lsh}. In the top panel of
Fig.~\ref{fig:balding}, we show the decay of the {horizon} magnetic flux
$\Phi_{\rm B}$ for all simulations. Overall, the decay times shown in
Fig.~\ref{fig:balding} are an order of magnitude shorter than those from the MHD
simulations of \cite{Bransgrove:2021heo} and \cite{Selvi:2024lsh}. This is
likely due to artificially high numerical resistivity (i.e., low spatial grid
resolution) compared to {those} studies. At higher resolution than we use, our
MHD solution will equally not be able to recover the correct collisionless
reconnection rate \citep{Sironi:2014jfa,Bransgrove:2021heo}. We therefore treat
our results mainly qualitatively, in that the BH pulsar forms and balds, and
defer quantitative conclusions to an analytical model discussed in Sec.
\ref{sec:striped wind}.

The magnetic flux decay timescale $\tau_\Phi$ is almost identical for both of
the inclined models, while the aligned model exhibits about 30\% faster decay
until $t-\tmerger \lesssim 6 {\rm ms}$. However, in a split monopole
magnetosphere of a stationary Kerr BH, the timescale $\tau_\Phi$ may not notably
depend on the current sheet inclination angle $\chi$ \citep{Selvi:2024lsh}. We
investigate the origin of the accelerated magnetic flux decay by examining
additional physical quantities, as follows.

Electromagnetic and gravitational perturbations around a BH can be analyzed by
means of the Newman-Penrose (NP) scalars
\citep{Teukolsky:1972my,Teukolsky:1973ha}
\begin{align}
    \psi_4 &= - C_{abcd} n^a \bar{m}^b n^c \bar{m}^d , \\
    \phi_2 &= F_{ab} \bar{m}^a n^b ,
\end{align}
where $C_{abcd}$ is the Weyl tensor, $F_{ab}$ is the electromagnetic field
tensor, and $(l^a, n^a, m^a, \bar{m}^a)$ are orthonormal null tetrads
\begin{align}
    l^a &= (t^a + r^a)/\sqrt{2} , \\
    n^a &= (t^a - r^a)/\sqrt{2} , \\
    m^a &= (\theta^a + i \phi^a)/\sqrt{2} .
\end{align}
We use the dominant $(l, m) = (2, 2)$ quadrupole mode of $\psi_4$ to monitor the
BH ringdown, and the $(l, m) = (1, 1)$ dipole mode of $\phi_2$ to monitor the
electromagnetic modulation in the magnetosphere.\footnote{The NP scalar $\psi_4$
corresponds to the outgoing gravitational radiation at null infinity. The
Maxwell NP scalar $\phi_2$ is proportional to the complex electric field
$E_\theta + i E_\phi$. In an axisymmetric background magnetic field, $E_\theta$
corresponds to the Alfv\'enic modes and $E_\phi$ to the fast magnetosonic
modes.}
We show the imaginary part of $\phi_2^{(l=1,m=1)}$ and $\psi_4^{(l=2,m=2)}$
(hereafter denoted simply as $\phi_2$ and $\psi_4$ for brevity) in the middle
and lower panels of Fig.~\ref{fig:balding}. In the following discussions, we
denote the exponential decay time scale of the NP scalar $\phi_2$ ($\psi_4$) as
$\tau^{\rm NP}_\phi$ ($\tau^{\rm NP}_\psi$).

We compute the quasi-normal mode (QNM) frequencies of the remnant BH for $(s, l,
m) = (-2, 2, 2)$ and $(s, l, m) = (-1, 1, 1)$ fundamental modes using the
\texttt{qnm} package \citep{Stein:2019mop}. The imaginary parts of the two QNM
frequencies are both around {$12\,r_g/c$}. From the real parts, we obtain the
oscillation period 0.94ms for $\phi_2$ and 0.59ms for $\psi_4$. The damped
sinusoidal oscillation of $\psi_4$, shown in the bottom panel of
Fig.~\ref{fig:balding}, agrees well with both real and imaginary parts of the
computed QNM frequency.

{\em Inclined magnetic field} ($\theta_B = 30^\circ, 60^\circ$) --- Both
simulations show $\tau^{\rm NP}_\phi = 31 r_g/c$, which is the same as {their}
magnetic flux decaying timescale $\tau_\Phi$. Periodic oscillations of $\phi_2$
for $t-\tmerger \gtrsim 1.5\,{\rm ms}$ are coming from the rotation of current
sheets (see Fig.~\ref{fig:cs}), which has a half-period of $2\pi/\Omega_H
\approx 1.8{\rm ms}$. From the fact that the measured decay timescales
$\tau_\Phi$ and $\tau^{\rm NP}_\phi$ not only agree with each other but also
being disparate from the QNM frequency, we deduce that $\tau_\Phi = 31 r_g/c$ is
the the flux decay timescale of the BH pulsar due to magnetic reconnection in
our setup, and the time decay of $\phi_2$ is simply a consequence of the
declining magnetic field strength.

{\em Aligned magnetic field} ($\theta_B = 0^\circ$) --- In the early phase of
the ringdown ($t-\tmerger \leq 5{\rm ms}$), $\phi_2$ exhibits a rapid decay with
$\tau^{\rm NP}_\phi\approx \tau^{\rm NP}_\psi$. The period of the oscillations
in $\phi_2$, which lasts about 2.5 cycles, is measured to be 0.9ms and shows a
good agreement with the QNM frequency (0.94ms). This indicates that the
evolution of the post-merger magnetosphere is dominated by the ringdown of the
BH, which rapidly sheds off magnetic fluxes from the horizon. We show $\phi_2$
in the meridional ($xz$) plane in Fig.~\ref{fig:ringdown magnetosphere}. The
magnetic flux shedding driven by QNMs induces episodes of quasi-periodic
modulations in the magnetosphere, which leads to a more rapid reconnection of
the field lines on the equatorial plane. The same process has been also observed
by \cite{Most:2024qgc} for the gravitational collapse of a NS with its spin axis
aligned with the magnetic moment.
On the other hand, both $\tau_\Phi$ and $\tau^{\rm NP}_\phi$ are slowed down to
$\approx 31 r_g/c$ later in $t-\tmerger \geq 6{\rm ms}$, implying that the
balding process of the BH begins to be affected more by resistivity. The
magnetic flux shedding by QNMs becomes subdominant as gravitational
perturbations fade out, then the flux decay is governed by magnetic reconnection
afterwards.
We caution that this observed behavior may change at higher numerical
resolutions, which will exhibit a better scale separation between plasma and
gravitational effects.

\subsection{Striped wind}
\label{sec:striped wind}

\begin{figure*}
    \centering
    \includegraphics[width=\linewidth]{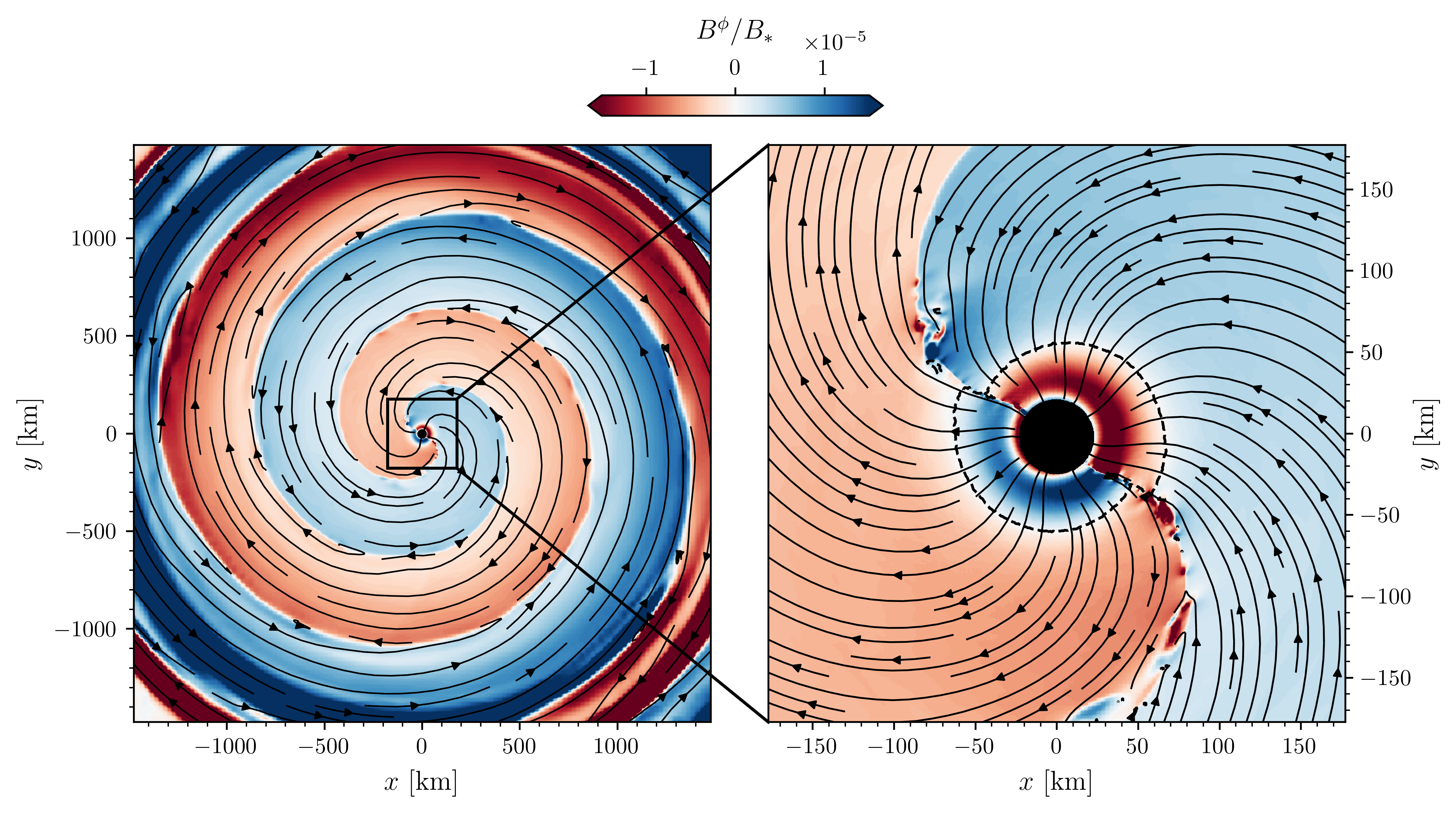}
    \caption{Pulsar-like striped wind from the remnant black hole at
        $t-\tmerger={\rm 7.0 ms}$ from $\theta_B = 30^\circ$ simulation. We show
        the toroidal magnetic field $B^\phi$ with the magnetic field lines on
        the equatorial ($xy$) plane in both panels. The remnant black hole is
        shown with a black circle and spinning counter clockwise in this figure.
        In the right panel, we show the stagnation surface ($u^r = 0$) with a
        black dashed line.}
    \label{fig:striped wind}
\end{figure*}

\begin{figure}
    \centering
    \includegraphics[width=\linewidth]{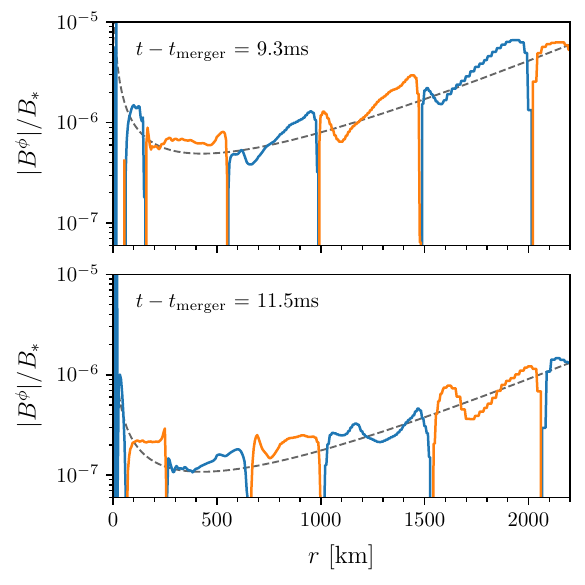}
    \caption{Toroidal magnetic field of the striped wind $|B^\phi(r)|$ on the
        equatorial plane along the $\hat{x}$ axis. Alternating signs
        (polaritires) of $B^\phi$ in each stripes are denoted with different
        colors. The dashed line shows the fit with Eq.~\eqref{eq:bphi}.}
    \label{fig:bphi-fitting}
\end{figure}

The rotation of an inclined split-monopole magnetosphere on a BH leads to a
striped wind \citep{Selvi:2024lsh} which appears to be similar to those from
oblique pulsars
\cite[e.g.][]{Michel:1982fj,Petri:2012cs,Tchekhovskoy:2012hm,Cerutti:2017hou},
albeit without the presence of a closed zone. We illustrate this in
Fig.~\ref{fig:striped wind} for the $\theta_B = 30^\circ$ simulation, where the
sign change in the toroidal magnetic field is clearly visible. Different from a
stationary pulsar solution with $B^\phi \sim 1/r$
\citep{Michel:1982fj,Bogovalov:1999vg}, the magnetic field here decays over
time. While this will not affect the geometry of the striped wind, its amplitude
will naturally become a function of retarded time, $B^\phi =B^\phi \left(r,
\theta, t- r/c\right)$, as can be seen from the left panel of
Fig.~\ref{fig:striped wind}.

A stagnation surface at which $u^r = 0$, separating the inflow and outflow
region of the plasma, appears in the vicinity of the BH
\citep{Bransgrove:2021heo}; we show it on the right panel of
Fig.~\ref{fig:striped wind}. Interestingly, the stagnation surface is
discontinuous across the rotating current sheet, with its radius being greater
at the upstream of the current sheet (trailing part of the striped wind),
appearing as a half-split spheroid with an offset along the current sheet.

We also find that the rotation angular velocity of the magnetic field lines is
slower (faster) at the upstream (downstream) of the rotating current sheet,
exhibiting a symmetric deviation from $\Omega_F = \Omega_H/2$. We reserve a more
detailed analysis of these near-horizon dynamics of oblique BH pulsars for
future work.

An analytic model of the toroidal magnetic field $B^\phi$ in the wind can be
developed as follows. For a nearly force-free wind from a rotating
split-monopole, $B^\phi$ can be approximated as
\citep{Michel:1982fj,Bogovalov:1999vg,Tchekhovskoy:2015cda}
\begin{equation} \label{eq:bphi-steady-monopole}
    |B^\phi (r,\theta)| \approx 
    \frac{\Omega r \sin \theta}{c} |B_r|
        = \frac{\Omega}{c} \frac{B_* r_*^2 \sin \theta}{r},
\end{equation}
where $\Omega$ is the rotation angular velocity, $B_*$ is the surface magnetic
field strength, and $r_*$ is the radius of the rotator. For a BH pulsar, we can
replace the angular velocity $\Omega$ with $\Omega_F = \Omega_H/2$, the radius
$r_*$ with $r_H$, and the surface magnetic field $B_*$ with $B_H(t) = B_{H,0}
e^{-t/\tau_\Phi}$. The resulting extension of
Eq.~\eqref{eq:bphi-steady-monopole} for a BH pulsar is
\begin{equation} \label{eq:bphi}
    |B^\phi(r,\theta,t)| = \frac{\Omega_H}{2c}
        \frac{B_{H,0} r_H^2 e^{-(t-r/c)/\tau_\Phi} \sin \theta}{r},
\end{equation}
where $(t-r/c)$ accounts for a retarded time.

Fig.~\ref{fig:bphi-fitting} compares the $\theta_B = 30^\circ$ simulation data
with Eq.~\eqref{eq:bphi} on the equatorial plane, using $\tau_\Phi = 31r_g/c$
measured from the balding process (Sec.~\ref{sec:balding and ringdown}) and
shifting $t\to t-\tmerger$. Our approximate analytic model shows a good
agreement with the simulation result.
The value of $B_{H,0}$ fitted from the simulation data is $1.5\e{-2} B_*$,
revealing that the split-monopole BH pulsar inherits about 1\% of the magnetic
field strength from the companion NS.
A separate estimate from the BH magnetic flux $\Phi_B = 2\pi r_H^2 B_{H,0}$ (top
panel of Fig.~\ref{fig:balding}) yields almost the same value of $B_{H,0}$,
reassuring the validity of the analytic model Eq.~\eqref{eq:bphi} as well as the
measured value of $B_{H,0}$.

\subsection{Energetics}
\label{sec:bh pulsar energetics}

The wind from the BH pulsar is powered by the energy extracted from the remnant
BH through the Blandford-Znajek (BZ) process \citep{Blandford:1977ds}, leading
to a spin-down of the BH. The spin-down power of an aligned split-monopole
magnetosphere, to a leading order of the BH spin,\footnote{The relative
correction from the next order term $\propto (\Omega_H)^4$ is less than
$10^{-3}$ in our case. See \cite{Tchekhovskoy:2009ba} for the expansion formula
up to $(\Omega_H)^6$.} is given as
\citep{Tchekhovskoy:2009ba}
\begin{equation} \label{eq:power-bz}
    P_{\rm BZ} = \frac{(\Phi_B^2/4\pi) \Omega_H^2}{6\pi c} .
\end{equation}
The BZ power \eqref{eq:power-bz} can be written into a form more commonly used
in the pulsar literature
\begin{equation} \label{eq:power-pulsar form}
    L = \frac{2}{3c} \Omega_F^2 B_H^2 r_H^4 ,
\end{equation}
with $\Omega_F = \Omega_H/2$ and $\Phi_B = 2\pi r_H^2 B_H$.

The spin-down power (Eq.~\eqref{eq:power-bz} or \eqref{eq:power-pulsar form}) is
carried by the electromagnetic Poynting flux, which is not a direct observable.
It is the dissipation in the current sheets which converts the electromagnetic
field energy of the wind into kinetic energy of particles
and subsequent electromagnetic emissions \cite[e.g.][]{Philippov:2019qud}. In a
steady pulsar magnetosphere, about 
10--20 percent of the spin-down 
power can be dissipated within 10 light cylinder radii
{(e.g., \citealt{Parfrey:2011ta,Chen:2014dva,Philippov:2014mqa}, see also
\citet{Cerutti:2016ttn} for a review).}

Here we develop a toy model for the dissipation luminosity of a BH pulsar,
closely following the approach by \cite{Cerutti:2020iqg}. From here we will use
$t$ to denote the time after the formation of the split monopole i.e.
$(t-\tmerger) \to t$. The total dissipation luminosity is given by a volume
integral
\begin{equation} \label{eq:dissipation-volume-integral}
\begin{split}
    L_D & = \int (\mathbf{J}\cdot\mathbf{E}) \, r^2 \sin\theta dr d\theta d\phi \\
    & = \frac{c \beta_{\rm rec}}{\pi} \int (B^\phi)^2 \, r \sin \theta dr d\theta ,
\end{split}
\end{equation}
where $\beta_{\rm rec}$ is the dimensionless reconnection rate
\citep{Uzdensky:2012tf}. A primary difference of our toy model from that of
\cite{Cerutti:2020iqg} is the exponential damping term in the
Eq.~\eqref{eq:bphi} associated with the flux decay of the BH. Substituting the
expression \eqref{eq:bphi} into \eqref{eq:dissipation-volume-integral} and
performing angular integration,
\begin{equation} \label{eq:dissipation-radial-integral}
    L_D = \frac{2\beta_{\rm rec} L_0}{\pi} e^{-2t/\tau_\Phi}
     \int_{r_{\rm min}}^{r_{\rm max}} \frac{e^{2r/c \tau_\Phi}}{r} dr ,
\end{equation}
where $L_0 = (2/3c) \Omega_F^2 B_{H,0}^2 r_H^4$ is an instantaneous BZ power of
the BH pulsar at $t=0$. The upper and lower bounds of the integral in
Eq.~\eqref{eq:dissipation-radial-integral} correspond to the radial extent of
the striped wind, $r_{\rm min} = r_H$ and $r_{\rm max} \approx ct$, which gives
\begin{equation} \label{eq:dissipation-luminosity}
    L_D(t) = \frac{2\beta_{\rm rec} L_0}{\pi} e^{-2t/\tau_\Phi}
    \left[
        {\rm Ei}\left(\frac{2t}{\tau_\Phi}\right)
        - {\rm Ei}\left(\frac{2r_H}{c \tau_\Phi}\right)
        \right] ,
\end{equation}
where ${\rm Ei}(x) = \int^x_{-\infty}(e^t/t) dt$ is the exponential intergral.

\begin{figure}
    \centering
    \includegraphics[width=\linewidth]{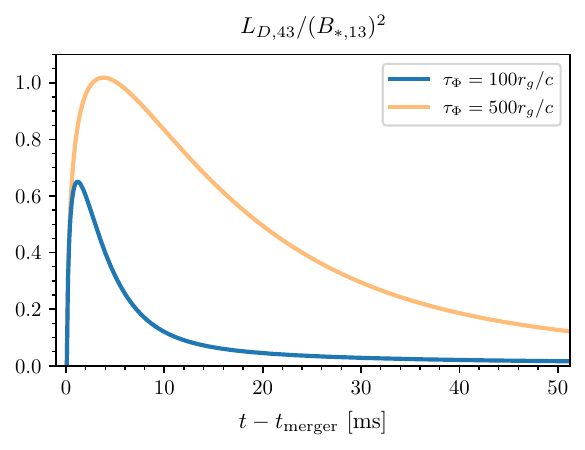}
    \caption{The dissipation luminosity from a BH pulsar $L_D(t)$ computed with
        an analytic model developed in Sec.~\ref{sec:bh pulsar energetics},
        normalized with $L_{D,43}\equiv L_D / (10^{43}{\rm erg \, s^{-1}})$ and
        $B_{*, 13}\equiv B_* / (10^{13}{\rm G})$. Due to a high (unphysical)
        numerical resistivity in our simulation, we construct the light curves
        using $\tau_\Phi = 100r_g/c$ (blue solid line) and $\tau_\Phi =
        500r_g/c$ (orange solid line) consistent with bounds from
        high-resolution kinetic simulations of \cite{Bransgrove:2021heo}.}
    \label{fig:dissipation-luminosity}
\end{figure}

We apply our toy model to the $\theta_B = 30^\circ$ simulation. The initial
spin-down power $L_0$ can be computed from the mass and spin of the remnant BH,
and using $B_{H,0}/B_* = 1.5\%$ fitted from the simulation result (see
Sec.~\ref{sec:striped wind}).\footnote{Note that this ratio $B_{H,0}/B_*$,
namely the portion of the magnetic flux that a nascent BH pulsar inherits from
the swallowed NS, can only be probed with a full numerical relativity merger
simulation as performed here.} The flux decay timescale $\tau_\Phi = 31 r_g/c$
from our simulation is dominated by unphysical numerical resistivity, therefore
we consider $\tau_\Phi = 100r_g/c$ and $\tau_\Phi = 500 r_g/c$ motivated from
the high-resolution (kinetic) simulations of \cite{Bransgrove:2021heo} as a more
realistic input for assessing light curves. The reconnection rate is fixed to
$\beta_{\rm rec} = 0.1$ from kinetic plasma simulations \citep{Sironi:2014jfa}.

In Fig.~\ref{fig:dissipation-luminosity}, we show the modelled dissipation
luminosity $L_D(t)$ scaled with the initial NS magnetic field strength. The time
curve of the dissipation luminosity exhibits a rapid rise to its peak value
within a few milliseconds, followed by exponential then power-law decay over
tens of milliseconds. A magnetar can power a burst with the luminosity $\sim
10^{47} {\rm erg \,s^{-1}}$, while a NS with $B_*\sim 10^{12} {\rm G}$ will emit
a relatively faint one with $\sim 10^{41} {\rm erg \,s^{-1}}$.
The exponential factor $e^{2r/c \tau_\Phi}$ in
Eq.~\eqref{eq:dissipation-radial-integral} suggests that the region $r\approx
r_{\rm max}$ is predominantly contributing to the total integral, implying the
forefront of the expanding striped wind with a thickness $\Delta r \approx c
\tau_\Phi$ is mainly powering the total dissipation luminosity.

The total dissipated energy $E = \int L_D(t) dt$ does not converge due to a
$t^{-1}$ asymptotic decay of $L_D(t)$. Realistically, dissipation in the current
sheets would introduce a faster decrease of $B^\phi$ in radius, and the decay of
$B_H(t)$ below a certain threshold can halt the pair production around the BH,
turning off the BH pulsar. Naively setting the end time of the burst as when
$L_D(t)$ drops down to $1/10$ of its peak value, the burst lasts about 15\,ms
(60\,ms) for $\tau_\Phi = 100 r_g/c$ ($500r_g/c$), with the average luminosity
$2.6\e{43}\, {\rm erg \, s^{-1}}$ ($4.2\e{43}\, {\rm erg \, s^{-1}}$) for $B_* =
10^{13}{\rm G}$.

\section{Electromagnetic transient}
\label{sec:em transient}

\subsection{Radio burst} 

The power dissipated in monster shocks at small radii is immediately radiated in
X-rays \citep{Beloborodov:2022pvn}. Later, when the shock expands to larger
radii, it can become a bright source of radio emission and emit a powerful {fast
radio burst (FRB)}. Magnetized shocks emit a radio precursor by the synchrotron
maser mechanism; it was initially proposed for termination shocks of pulsar
winds \citep{Hoshino1992,Lyubarsky:2014jta} and then for internal shocks in
magnetized $e^\pm$ outflows to explain repeating FRBs
\citep{Beloborodov:2017juh}.

Consider first the monster shock at small radii $r\sim
10^7$\textendash{}$10^8$\,cm. Kinetic plasma simulations of magnetized shocks
\citep{Sironi:2021wca,Vanthieghem:2024van} show precursor emission with
frequency $\omega_{\rm pre}\sim 3\tilde{\omega}_B=3e\tilde{B}/m_e c$, where
$\tilde{B}$ is the upstream magnetic field measured in the plasma rest frame,
$e$ is the elementary charge, and $m_e$ is the electron mass. $\tilde{B}$ is
reduced from the background value $B_{\rm bg}$ by the strong expansion of the
plasma ahead of the monster shock \citep{Beloborodov:2022pvn}: 
\begin{equation}
    \tilde{B}\approx \frac{\omega r}{2c\sigma_{\rm bg}}B_{\rm bg},
\end{equation}
where $\sigma_{\rm bg}=B_{\rm bg}^2/4\pi n_{\rm bg} m_e c^2$ is the background
magnetization parameter, and $\omega$ is the frequency of the magnetospheric
perturbation that led to shock formation (our simulation shows $\omega r/c\sim
10$). Density $n_{\rm bg}$ can be parameterized by multiplicity ${\cal M}\equiv
n_{\rm bg}/n_0$, where $n_0=\nabla\cdot\boldsymbol{E}/4\pi e\sim \Omega B_{\rm
bg}/2\pi e c $ is the minimum density required to support the magnetospheric
rotation with drift speed $\sim \Omega r$ \citep{Goldreich:1969sb}. 
This gives $\omega_{\rm pre}\sim (r\omega/c) {\cal M}\Omega \sim 10^4 {\cal
M}$\,rad/s.
This simple estimate is, however, deficient because it neglects the deceleration
of the upstream flow by strong radiative losses. Losses dramatically change the
radio precursor from monster shocks at small radii by increasing its frequency
and suppressing its power (Beloborodov, in preparation).

Powerful radio emission is expected from the relativistic shock when it expands
far into the $e^\pm$ outflow, reaching $r\sim 10^{13}$\textendash{}$10^{14}$\,cm
\citep{Beloborodov:2017juh,Beloborodov:2019wex}. Then, a fraction $\sim 10^{-4}$
of the blast wave power is expected to convert to radio waves, whose frequency
decreases with time (proportionally to the local $B$) and passes through the GHz
band, best for radio observations. In this paper, we do not follow the outflow
dynamics with shocks at large radii; however, this may become possible for
future MHD simulations. Our simulation shows that shocks launched from NS-BH
mergers are asymmetric, but not strongly collimated. Therefore, they can produce
FRBs observable for a broad range of line of sights. Note that no baryonic
ejecta are expected from BH swallowing a NS, so nothing should block the FRB
from observers.

\subsection{Gamma-ray burst}

The X-ray transient expected from the simulated merger is powered by dissipation
of magnetospheric energy. Two dissipation mechanisms are observed in the
simulation: shocks and magnetic reconnection in the split-monopole current sheet
around the BH after the merger. Dissipation occurs at small radii, which
correspond to a large compactness parameter $\ell= \sigma_{\rm T}L/r m_ec^3$,
where $L$ is the dissipation power and $\sigma_{\rm T}$ is the Thompson cross
section. Note that $L$ and $\ell$ scale as $B^2$. For a strongly magnetized NS,
e.g. with $B\sim 10^{14}\,{\rm G}$, the huge $\ell$ implies that the dissipated
energy becomes immediately thermalized. Thus, the merger ejects a hot
``fireball'' -- a thermalized, magnetically dominated $e^\pm$ outflow. As the
outflow expands to larger radii, it adiabatically cools, $e^\pm$ annihilate and
release a burst of quasi-thermal radiation similar to the GRB from the magnetar
collapse described in \cite{Most:2024qgc}.

An additional dissipation mechanism is expected to operate in the outflow at
large radii, and can add a nonthermal tail to the GRB spectrum. It is caused by
the striped structure of the outflow, similar to the striped winds from pulsars.
The stripes develop current sheets where magnetic reconnection gradually
dissipates the alternating magnetic flux
\citep{Lyubarsky:2000yn,Cerutti:2020iqg}. A similar mechanism was previously
proposed to operate in canonical GRBs \citep{Drenkhahn:2002ug}. It will release
energy after the outflow becomes optically thin (which happens quickly in the
baryon-free outflow from BH--NS merger). Therefore, it can generate energetic
particles, emitting a nonthermal component of the GRB.

\section{Conclusions} \label{sec:conclusion}

We have presented a detailed numerical investigation into the magnetospheric
dynamics of BH--NS mergers without tidal disruption. Using GRMHD simulations
capable of probing the near force-free limit, we identify two mechanisms for
generating an electromagnetic transient.

First, we observe that fast magnetosonic waves
are launched into the magnetosphere of the NS before it plunges into the BH.
These waves, as expanding outward with almost the speed of light, develop into
monster shocks due to a more rapidly decaying ambient magnetic field
\citep{Beloborodov:2022pvn}.
The full MHD simulation is essential for tracking this effect, so it could not
be captured by earlier vacuum or force-free simulations. The launched shocks are
expected to emit a bright radio transient when they expand to large radii.

When the BH swallows the NS together with its magnetic dipole moment, its
external magnetosphere quickly rearranges itself 
into a split-monopole configuration with a large-scale current sheet. Then, the
BH gradually loses the acquired ``magnetic hair.'' This balding is assisted by
magnetic reconnection and gravitational effects (QNMs).
The relative importance of these two processes varies over time and depends on
the misalignment between the magnetic dipole moment and the BH spin.
The split monopole is dragged into rotation by the BH and forms a transient BH
pulsar which can power a post-merger EM signal in the X-ray and $\gamma$-ray
band. 

The monster shocks and the balding BH pulsar were previously studied in
symmetric setups 
with a single compact object
\citep{Bransgrove:2021heo,Beloborodov:2022pvn,Most:2024qgc,Selvi:2024lsh}. Our
ab-initio simulations demonstrate how both phenomena naturally occur in the
complex dynamical spacetime of the BH--NS merger.

The binary parameters considered in our work are representative of the BH--NS
mergers detected to date \citep{LIGOScientific:2024elc}, implying that shock
formation from magnetosonic waves and the emergence of a BH pulsar could be a
common outcome for the BH--NS populations observable with ground-based GW
detectors such as the LIGO/Virgo/KAGRA network.

\section*{Acknowledgments}

The authors are grateful to Ashley Bransgrove, Koushik Chatterjee, Alexander
Chernoglazov, Amir Levinson, Keefe Mitman, Alexander Philippov, Eliot Quataert,
Sebastiaan Selvi, Lorenzo Sironi, Anatoly Spitkovsky, Alexander Tchekhovskoy,
Saul Teukolsky, Christopher Thompson, and Yici Zhong for insightful comments and
discussions.
YK acknowledges support by the Sherman Fairchild Foundation and by NSF Grants
No. PHY-2309211, No. PHY-2309231, and No. OAC-2209656 at Caltech.
ERM acknowledges support by the National Science Foundation under grants No.
PHY-2309210 and AST-2307394, and from NASA's ATP program under grant
80NSSC24K1229.
AMB acknowledges support by NASA grants 80NSSC24K1229 and 21-ATP21-0056, and
Simons Foundation grant No. 446228.
BR acknowledges support by the Natural Sciences \& Engineering Research Council
of Canada (NSERC), the Canadian Space Agency (23JWGO2A01), and by a grant from
the Simons Foundation (MP-SCMPS-00001470). BR acknowledges a guest researcher
position at the Flatiron Institute, supported by the Simons Foundation. 

Simulations were performed on the NSF Frontera supercomputer at the Texas
Advanced Computing Center under grant AST21006, and on the Delta cluster at the
National Center for Supercomputing Applications (NCSA) through allocation
PHY210074 from the Advanced Cyberinfrastructure Coordination Ecosystem: Services
\& Support (ACCESS) program, which is supported by National Science Foundation
grants \#2138259, \#2138286, \#2138307, \#2137603, and \#2138296.

\software{
    EinsteinToolkit \citep{Loffler:2011ay},
    Frankfurt/IllinoisGRMHD \citep{Most:2019kfe,Etienne:2015cea}
    FUKA \citep{Papenfort:2021hod},
    Kadath \citep{Grandclement:2009ju},
    AHFinderDirect \citep{Thornburg:2003sf},
    kuibit \citep{kuibit},
    matplotlib \citep{matplotlib},
    numpy \citep{numpy},
    scipy \citep{scipy},
    qnm \citep{Stein:2019mop}
}

\bibliography{references}

\end{document}